# Performance evaluation of non-uniform sensor spacing in a linear array configuration for MUSIC algorithm

Pradeep D., Sumit Saraogi, Palanisamy P., Kalyanasundaram N.

Department of Electronics & Communication Engineering, National Institute of Technology Tiruchirappalli -15, India.

**ABSTRACT**

*In this paper, the performance of non-uniform spacing of sensors is evaluated for the MUSIC algorithm which estimates the direction of arrival (DOA) of a narrowband plane wave impinging on an array of sensors. Unlike uniform sensor spacing arrangement, where sensors are equidistant (equal to half the wavelength), we consider non-uniform spacing for the arrangement of sensors, where the distance between consecutive sensors increases progressively. We observe that the non-uniform sensor spacing configuration (with lesser number of sensors) can provide similar or better accuracy in DOA estimation compared to uniform sensor spacing configuration despite more number of sensors at identical array length.*

## 1. INTRODUCTION

The MUSIC algorithm [1] is a MUltiple SIgnal Classification technique based on exploiting the Eigen structure of the input covariance matrix. MUSIC is a signal parameter estimation algorithm which provides information about the number of incident signals, Direction-of-arrival (DOA) of each signal, strengths and cross- correlation between incident signals, noise powers, etc. While MUSIC provides very high resolution, it requires very precise array calibration.

The array length is defined as the distance between the first and the last sensor in the array configuration. The sensor arrangement in the array can be of two possible types - Uniform sensor array spacing (Fig.1), where the distance between the consecutive array sensor elements is constant and equal to half the wavelength and Non-uniform sensor spacing (Fig.2), where the sensor spacing is not constant and the distance between the consecutive sensors increases progressively. We evaluate the performance of both these sensor spacing arrangements for DOA estimation using MUSIC algorithm while keeping the array length identical.

For uniform sensor spacing, the array length is multiple of half the wavelength, whereas in non-uniform sensor spacing, it is defined as the distance between the first and the last sensor in the configuration. In all the simulation results presented in this paper we have considered the array length to be identical for comparison between uniform and non-uniform sensor spacing cases.

We have considered the mathematical expression [2, 3] for the perturbation or error in the estimation of DOA using MUSIC algorithm. We also calculate this perturbation for both uniform sensor spacing and non-uniform sensor spacing cases via simulations in MATLAB. We observe that the error in DOA estimation for non-uniform sensor spacing is lower or comparable to the uniform sensor spacing.

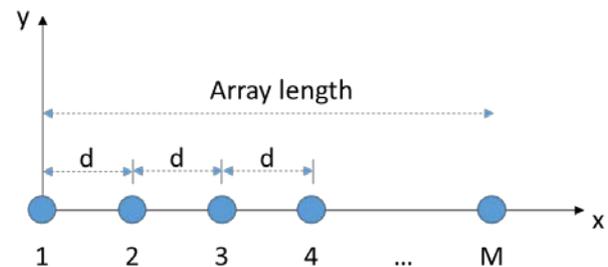

Fig.1 Uniform Sensor Spacing

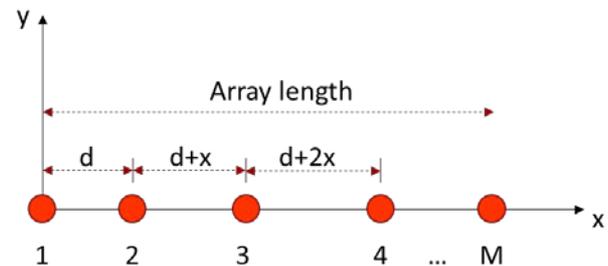

Fig.2 Non-uniform Sensor Spacing

The terminologies used in this paper are as follows: Superscript T and H denote transpose and conjugate transpose respectively and upper/lower case bold letter denotes matrix/vector.

## 2. ARRAY SIGNAL MODEL

Consider an array composed of M sensors located in x-y plane, and assume that D (<M) narrow-band signals, centered around a known frequency, say $\omega_c$, impinging on the array from distinct directions $\theta_1, \theta_2, ..., \theta_D$. For simplicity assume the sources and sensors are located in the same plane and the sources are in the far field of the array. In this case, the only parameter that characterizes the source location is its direction of arrival angle '$\theta$'.

Using complex envelope representation, the M×1 vector received by the array can be expressed, in matrix notation, as follows

$$x(t) = A(\theta) s(t) + n(t) \qquad (1)$$

where, $x(t) \triangleq [x_1(t)\ x_2(t)\ \cdots x_M(t)]^T$

$s(t) \triangleq [s_1(t)\ s_2(t)\ \cdots s_D(t)]^T$

$n(t) \triangleq [n_1(t)\ n_2(t)\ \cdots n_M(t)]^T$

and $A(\theta)$ is M×D matrix of the steering vectors :

$$A(\theta) \triangleq [a(\theta_1)\ a(\theta_2)\ \cdots\ a(\theta_D)]$$

The steering vector of the array towards the direction $\theta$ is given as equation (2)

$$a(\theta) = \left[ e^{j\frac{2\pi}{\lambda} p_1 \cos\theta + q_1 \sin\theta} \cdots e^{j\frac{2\pi}{\lambda} p_M \cos\theta + q_M \sin\theta} \right]^T$$

$s_i(t)$ – the signal of the i$^{th}$ source
$n_i(t)$ – the noise at i$^{th}$ sensor
$(p_i, q_i)$ – the coordinate of the i$^{th}$ sensor.

The DOA estimation problem is to estimate the locations $\theta_1, \theta_2, \ldots, \theta_D$ of the source from the N samples (snapshots) of the array:
$x(t_1), x(t_2), \ldots x(t_N)$.

## 3. MUSIC

The MUSIC algorithm is a MUltiple SIgnal Classification technique based on exploiting the Eigen structure of the covariance matrix of the array input vector x(t). MUSIC is a signal parameter estimation algorithm which provides information about the number of incident signals, Direction-of-arrival of each signal, strengths and cross-correlation between incident signals, noise powers etc.

In terms of the above signal model the covariance matrix $R = E[x(t) x^H(t)]$ can be expressed in terms its eigenvalues: $e_1, e_2, \ldots, e_M$ and eigen vectors: $v_1, v_2, \ldots, v_M$ as $R = \Phi \Lambda \Phi^H$

where $\Lambda = \text{diag}(e_1, e_2, \ldots, e_M)$ and $\Phi = [v_1, v_2, \ldots, v_M]$

We assume that the eigenvalues are in the order of decreasing size as: $e_1 \geq e_2 \geq \cdots \geq e_D > e_{D+1} = e_M$.

The D larger eigenvalues are called signal level eigenvalues and the others are referred as the noise level eigenvalues.

The noise level eigen vectors are orthogonal to the steering vectors associated with the incident plane wave, that is: $\langle a(\theta), v_i \rangle = 0$, for $i = D+1, D+2, \ldots, M$ where $\langle\ \rangle$ denotes the inner product.

This is the essential observation of the MUSIC approach. It means that one can estimate the steering vectors associated with the received signals by finding the steering vectors which are most closely orthogonal to the eigenvectors associated with the eigenvalues of R that are approximately equal to noise level eigenvalues.

This analysis shows that the eigenvectors of the covariance matrix R belong to either of the two orthogonal subspaces, called the signal subspace and the noise subspace. The steering vectors corresponding to the Directions-Of-Arrival lie in the signal subspace and are hence orthogonal to the noise subspace. By searching through all possible array steering vectors to find those which are perpendicular to the space spanned by the non-principal eigenvectors, the DOAs can be determined.

To search the noise subspace, we form a matrix containing the noise eigenvectors:

$$V_n = [v_D\ v_{D+1}\ \ldots\ v_M] \qquad (3)$$

Since the steering vectors corresponding to signal components are orthogonal to the noise subspace eigenvectors,

$$a^H(\theta) V_n V_n^H a(\theta) = 0$$

for all $\theta$ corresponding to the DOA of a multipath component. Then the DOAs of the multipath incident signals can be estimated by locating the peaks of a MUSIC spatial spectrum given by

$$\psi(\theta) = \frac{1}{a^H(\theta) V_n V_n^H a(\theta)} \qquad (4)$$

## 4. PERTURBATION OF DOA

### 4.1 Mathematical derivation of our proposal

For MUSIC algorithm, the null spectrum function is given by

$$f(\theta, V_n) = \{\psi(\theta)\}^{-1} = a^H(\theta) V_n V_n^H a(\theta) \quad (5)$$

In noisy environment, estimated Direction-of-Arrival of the signal is denoted as perturbation from the actual Angle-of-Arrival $\qquad \hat{\theta} = \theta + \Delta\theta \qquad (6)$

where $\Delta\theta$ is the perturbation of the Angle-of-Arrival of the signal. So

$$f(\hat{\theta}, V_n) = f(\theta + \Delta\theta, V_n) \quad (7)$$

Using Taylor series expansion the equation (7) can be expressed as equation (8) given below

$$f(\hat{\theta}, V_n) = \sum_{k=0}^{\infty} (\Delta\theta)^k \frac{f^{(k)}(\theta, V_n)}{k!}$$

where, $f^{(k)}(\theta, V_n) \triangleq \frac{\partial^k}{\partial \theta^k} f(\theta, V_n)$.

The eq. (8) can be rearranged as eq. (9) as below

$$\frac{\partial}{\partial \theta} f(\hat{\theta}, V_n) = \sum_{k=1}^{\infty} (\Delta\theta)^{k-1} \frac{f^{(k)}(\theta, V_n)}{k!}$$

where $\frac{\partial}{\partial \theta} f(\hat{\theta}, V_n) \triangleq \frac{f(\theta + \Delta\theta, V_n) - f(\theta, V_n)}{\Delta\theta}$

Neglecting higher powers of $\Delta\theta$ in eq (9), we get equation (10) as given below

$$\frac{\partial}{\partial \theta} f(\hat{\theta}, V_n) = \frac{f^{(1)}(\theta, V_n)}{1!} + \Delta\theta \frac{f^{(2)}(\theta, V_n)}{2!}$$

The LHS in equation (10) is zero since $f(\hat{\theta}, V_n)$ attains minima at $\theta$. Therefore we get eq (11)

$$\Delta\theta = -\frac{2\,f^{(1)}(\theta, V_n)}{f^{(2)}(\theta, V_n)}$$

Now, $f^{(1)}(\theta, V_n)$ can be derived from eq. (5) as

$$f^{(1)}(\theta, V_n) = \frac{\partial}{\partial \theta}[a^H(\theta) V_n V_n^H a(\theta)] \quad (12)$$

$$f^{(1)}(\theta, V_n) = d1 + d2$$

where, $d_1 = \frac{\partial}{\partial \theta}[a^H(\theta)] V_n V_n^H a(\theta)$ and

$$d_2 = a^H(\theta) V_n V_n^H \frac{\partial}{\partial \theta}[a(\theta)]$$

We can show that $d_2 = d_1^H$. Therefore the equation (12) can be written as

$$f^{(1)}(\theta, V_n) = 2 \cdot Re\left\{a^H(\theta) V_n V_n^H \frac{\partial}{\partial \theta}[a(\theta)]\right\} \quad (13)$$

where $Re\{z\}$ denotes real part z. And get eq. (14) as

$$f^{(2)}(\theta, V_n) = \frac{\partial}{\partial \theta} f^{(1)}(\theta, V_n) = T_1 + 2 \cdot T_2 + T_3$$

where, $T_1 \stackrel{\text{def}}{=} \frac{\partial^2}{\partial \theta^2}[a^H(\theta)] V_n V_n^H a(\theta)$

$$T_2 = \frac{\partial}{\partial \theta}[a^H(\theta)] V_n V_n^H \frac{\partial}{\partial \theta}[a(\theta)]$$

$$T_3 = a^H(\theta) V_n V_n^H \frac{\partial^2}{\partial \theta^2}[a(\theta)]$$

It can be shown that $T_3 = T_1^H$. So equation (14) can be written as below

$$f^{(2)}(\theta, V_n) = 2 \cdot [T_2 + Re(T_3)] \quad (15)$$

Substituting equation (13) and (15) in (11) we get the perturbation of the Angle-of-Arrival as

$$\Delta\theta = \frac{-2\,Re\left\{a^H(\theta) V_n V_n^H \frac{\partial[a(\theta)]}{\partial \theta}\right\}}{\frac{\partial[a^H(\theta)]}{\partial \theta} V_n V_n^H \frac{\partial[a(\theta)]}{\partial \theta} + Re\left\{a^H(\theta) V_n V_n^H \frac{\partial^2[a(\theta)]}{\partial \theta^2}\right\}}$$

## 5. SIMULATION AND OUTPUT

In this section we compare RMSE (root mean square error) values computed by the theoretical expression derived in section 4 with the simulated RMSE values for the MUSIC algorithm.

$$RMSE = \sqrt{\sum_{i=1}^{L}(\theta_i - \hat{\theta}_i)^2}$$

Let RMSE be the root mean square error of the direction estimates, where $\hat{\theta}_i$ is the estimate value of $\theta_i$ and L is the number of trials.

The following parameters are fixed for all the results presented in this section:

Number of source, D=1; Number of Snapshots, N = 200; Center frequency, $\omega_c = 850$ MHz, Resolution = 0.01 degree and Number of trials, L = 100.

Using MATLAB, we calculate the RMSE in DOA estimation using the theoretically derived expression above. Then we compare this result with the RMSE in DOA estimation using simulations of the MUSIC algorithm. We plot a graph of RMSE in DOA estimation of theoretical and simulation results for different numbers of sensors considering both uniform and non-uniform sensor spacing for SNRs: -5dB, 0dB, 5dB and 10dB. We observe that the simulation results and calculated errors vary similarly and the simulation results are higher than the calculated theoretical values by a factor of around 28. This is due to the omission of higher order terms in the Taylor series expansion in equation (8). The graph in Fig.3.1 and Fig.3.2 shows, respectively, theoretical and simulated RMSE, for different sensors with uniform and non-uniform spacing, for various SNR values for an array length of $10(\lambda/2)$ and source direction: $\theta = 60°$. Similarly, Fig.3.3 and Fig.3.4 summarizes the result for an array length of $11(\lambda/2)$ and source direction $\theta = 50°$. These results are also tabulated in Table 1 and Table 2.

**Table 1** RMSE in DOA estimation using MUSIC algorithm for an array length of $10(\lambda/2)$ and a source direction $\theta = 60°$

| SNR (dB) | RMSE values | | | | | |
|---|---|---|---|---|---|---|
| | Theoretical result x10<sup>-4</sup> | | | Simulation result x10<sup>-3</sup> | | |
| | M=5 | M=8 | M=11 | M=5 | M=8 | M=11 |
| -5 | 2.5 | 3.1 | 3.0 | 72.9 | 88.1 | 85.9 |
| 0 | 2.5 | 3.0 | 3.0 | 72.5 | 85.3 | 85.9 |
| 5 | 3.0 | 3.3 | 3.3 | 86.7 | 93.8 | 93.5 |
| 10 | 3.0 | 3.3 | 3.2 | 71.6 | 80.5 | 91.4 |

**Table 2** RMSE in DOA estimation using MUSIC algorithm for an array length of $11(\lambda/2)$ and a source direction $\theta = 50°$

| SNR (dB) | RMSE values | | | | | |
|---|---|---|---|---|---|---|
| | Theoretical result x10<sup>-4</sup> | | | Simulation result x10<sup>-3</sup> | | |
| | M=6 | M=9 | M=12 | M=6 | M=9 | M=12 |
| -5 | 2.1 | 2.7 | 3.6 | 61.4 | 78.0 | 103.7 |
| 0 | 2.0 | 2.5 | 3.7 | 56.4 | 71.6 | 105.8 |
| 5 | 1.9 | 2.6 | 3.0 | 54.6 | 73.3 | 85.7 |
| 10 | 2.3 | 2.7 | 3.0 | 65.4 | 79.2 | 87.1 |

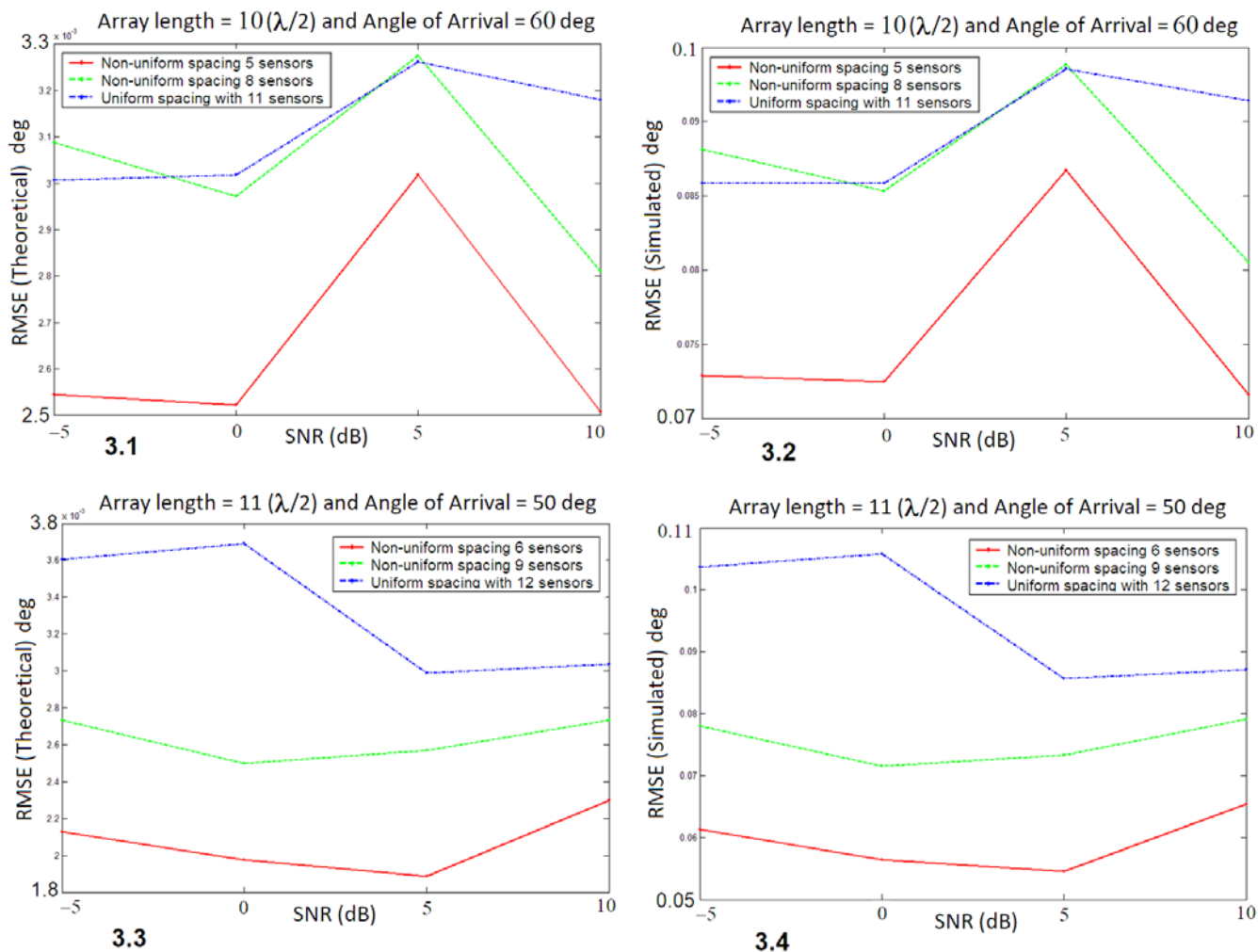

**Figure 3**. Comparison of RMSE error in DOA estimation for non-uniform sensor spacing (red and green) and uniform sensor spacing (blue) using theoretical (subpanel 3.1 and 3.3) and simulation results (subpanel 3.2 and 3.4) for different array lengths and angles of arrival. One can infer that the non-uniform sensor spacing (with lesser number of sensors) has either lower or similar accuracy in DOA estimation when compared to uniform sensor spacing of equal array length.

## 6. CONCLUSION

This paper provides a fair idea of the relative tradeoffs involved in the non-uniform sensor spacing using the derived mathematical expression for perturbation in the DOA estimation for MUSIC algorithm. Further, the accuracy of this mathematical expression is verified by simulation results of RMSE in DOA estimation for MUSIC algorithm. For a given array length, the non- uniform sensor spacing has lesser number of sensors in comparison to the uniform sensor spacing and we observe that few particular cases of non-uniform sensor spacing arrangement provide comparable or even better RMSE performance than the uniform sensor spacing arrangement.

## Reference

[1] R. Schmidt, "Multiple emitter location and signal parameter estimation," in IEEE Transactions on Antennas and Propagation, vol. 34, no. 3, pp. 276-280, March 1986, doi: 10.1109/TAP.1986.1143830.

[2] F. Li and R. J. Vaccaro, "Analysis of Min-Norm and MUSIC with arbitrary array geometry," in IEEE Transactions on Aerospace and Electronic Systems, vol. 26, no. 6, pp. 976-985, Nov. 1990, doi: 10.1109/7.62249.

[3] F. Li, H. Liu and R. J. Vaccaro, "Performance analysis for DOA estimation algorithms: unification, simplification, and observations," in IEEE Transactions on Aerospace and Electronic Systems, vol. 29, no. 4, pp. 1170-1184, Oct. 1993, doi: 10.1109/7.259520.

[4] Harry L. Van Trees, Optimum Array Processing: Part IV of Detection, Estimation and Modulation Theory (Wiley-Interscience, 2002).

[5] Joseph C. Liberti Jr & Theodore S. Rappaport, Smart Antennas for Wireless Communication-IS-95 and Third Generation CDMA Applications